\documentclass{article}
\usepackage{spconf,amsmath,graphicx,amssymb,booktabs,bm}

\title{ Masks Fusion with Multi-Target Learning For Speech Enhancement}
\name{Liangchen Zhou$^1$, Wenbin Jiang$^2$, Jingyan Xu$^1$, Fei Wen$^1$, Peilin Liu$^1$}
\address{
	$^1$Department of Electronic Engineering, Shanghai Jiao Tong University, Shanghai, China\\
	$^2$Department of Computer Science and Engineering, Shanghai Jiao Tong University, Shanghai, China}
\begin{document}
%\ninept
%
\maketitle
\begin{abstract}
Recently, deep neural network (DNN) based time-frequency (T-F) mask estimation
has shown remarkable effectiveness for speech enhancement.
Typically, a single T-F mask is first estimated based on DNN and
then used to mask the spectrogram of noisy speech in an order to suppress the noise.
This work proposes a multi-mask fusion method for speech enhancement.
It simultaneously estimates two complementary masks,
e.g., ideal ratio mask (IRM) and target binary mask (TBM),
and then fuse them to obtain a refined mask for speech enhancement.
The advantage of the new method is twofold.
First, simultaneously estimating multiple complementary masks brings benefit endowed by multi-target learning.
Second, multi-mask fusion can exploit the complementarity of multiple masks to boost the performance of speech enhancement.
Experimental results show that the proposed method can achieve significant PESQ improvement and reduce the recognition error rate of back-end over traditional masking-based methods.
Code is available at https://github.com/lc-zhou/mask-fusion.
\end{abstract}
\section{Introduction}
\label{sec:intro}

Speech enhancement has long been a fundamental and important problem in speech signal processing,
of which the goal is to suppress the interfering noise of observed noisy speech to improve the speech intelligibility and perceptual quality.
It has wide applications such as mobile telecommunication, automatic speech recognition (ASR), hearing prosthesis, and speech interaction, to name just a few \cite{8369155}.

\begin{figure}[t]
	\centering
	\includegraphics[width=\linewidth]{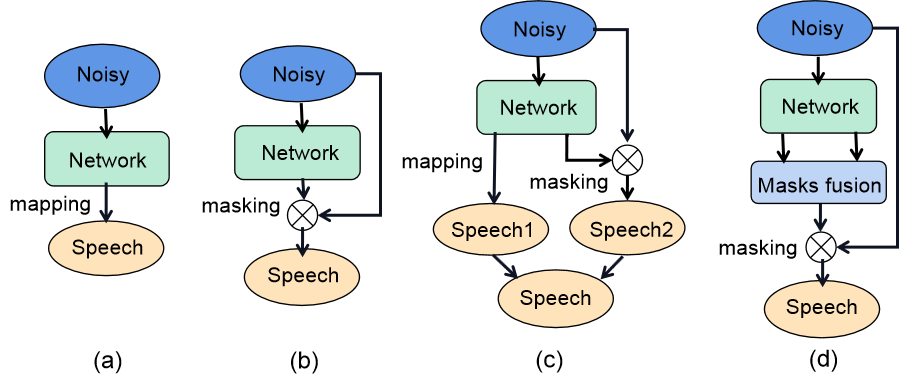}
	\caption{Illustration of different DNN-based speech enhancement methods. (a) Direct speech spectrogram estimation. (b) Masking-based method estimates a T-F mask first. (c) Simultaneous estimation of speech and a mask using DNN and fuse them for enhancement. (d) Our method simultaneously estimates two complementary masks and fuse them.}
	\label{fig:methods of se}
\end{figure}

Recently, due in part to the roaring success of deep learning, the research of deep neural
network (DNN) based speech enhancement has attracted much
attention and achieved much progress.
Generally, existing DNN based methods can be conveniently classified into three categories.
The first directly maps noisy speech to the target speech using DNN \cite{xu2014regression,han2015learning}, as illustrated in Figure \ref{fig:methods of se}(a).
The second is time-frequency (T-F) masking based \cite{williamson2015complex,narayanan2013ideal},
which estimates a mask using DNN and then applies it in the T-F domain of noisy speech to suppress the noise, as illustrated in Figure \ref{fig:methods of se}(b).
The third is a combination of the two formers,
which estimates the target speech and a mask simultaneously,
and fuse them in an attempt to achieve better performance \cite{7895577,weninger2015speech,ge2019environment,9054661},
as illustrated in Figure \ref{fig:methods of se}(c).
Roughly speaking, considering both the effectiveness and efficiency, T-F masking based methods are more prevalent.

Besides, there exist some works using multi-target architectures to exploit additional information benefiting denoising,
such as SNR-aware \cite{fu2016snr}, speaker-aware feature \cite{9053214}, spatial-aware masks \cite{jiang2018robust} and phase of noisy wav \cite{8691424}.
It has been shown that exploiting such information can improve the performance for speech enhancement.
Meanwhile, some other works design more complex neural network for denoising, e.g., \cite{9054615}.
These methods improve the performance of speech enhancement to a certain degree at the cost of increased complexity of models.

This paper proposes a multi-mask fusion method for speech enhancement.
It simultaneously estimates two complementary masks (see Figure \ref{fig:methods of se}(d)), ideal ratio mask (IRM) and target binary mask (TBM),
and then fuse them to obtain a refined mask for speech enhancement.
The motivation of the proposed method is twofold.
First, simultaneously estimation of multiple complementary masks can be expected to have benefit from multi-target learning.
Second, fusion of multiple masks can be expected to achieve performance improvement by incorporating complementary information from multiple masks.
Moreover, the proposed method does not incur much increase of network complexity,
and hence can well preserve the computational efficiency of T-F masking-based methods.
Experimental results demonstrate that the proposed method can achieve significant improvement in PESQ
compared with traditional masking-based methods.

Furthermore, when incorporated with sophisticated perceptual loss,
the proposed fusion scheme can yield further improvement
and outperforms existing state-of-the-art perceptual loss based method \cite{fu2019metricGAN}.
Recently, training the model of speech enhancement by a human-perception-related loss function has shown high effectiveness \cite{ fu2019metricGAN, martin2018deep, 8902088, kolbaek2020loss}.
Typically, such methods use an approximated PESQ or short-time objective intelligibility (STOI) loss function to train the network instead of the mean squared error (MSE) loss.

The rest of this paper is organized as follows.
Section 2 briefly introduces the signal model and background.
Section 3 presents the proposed mask fusion method.
Section 4 provides experiments, and Section 5 summarizes this paper.

\section{Preliminaries}
\label{sec:format}

An observed noisy speech signal ${\mathbf{y}}\in \mathbb{R}^{T}$  with $T$ sampling points
can be modeled as a mixture of a clean target speech $\mathbf{x}$ and noise ${\mathbf{n}}$ as
\begin{equation}\label{eq1_signal_model}
{\mathbf{y}} = {\mathbf{x}} + {\mathbf{n}}.
\end{equation}
The goal of speech enhancement is to recover ${\mathbf{x}}$ from ${\mathbf{y}}$.
In practice, ${\mathbf{x}}$ may consist of speech from multiple speakers.

Recently, DNN based methods has substantially advanced the performance of speech enhancement.
Especially, masking based methods have been prevalent due to their effectiveness and efficiency,
which typically estimate a T-F mask using DNN and then apply it to T-F domain spectrogram of noisy speech to suppress the noise.
Specifically, the time-domain observation ${\mathbf{y}}$ is transformed into T-F domain by short-time Fourier transform (STFT),
denoted by $F: \mathbb{R}^{T}\rightarrow \mathbb{C}^{M \times L} $, where ${M}$ and ${L}$ are the numbers of frames and frequency bins, respectively.
Generally, masking-based speech enhancement can be expressed as
\begin{equation}\label{Eq1}
\hat{{\mathbf{x}}}=F^{\dagger}\big(G(F({\mathbf{y}});{\mathbf{\Theta}}) \odot F({\mathbf{y}})\big),
\end{equation}
where $\hat{\mathbf{x}}$ is the estimation of ${\mathbf{x}}$, $ F^{\dagger} $ is the inverse-STFT,
$\odot$ is the element-wise Hadamard product, $G$ is the DNN mapping for T-F mask estimation with parameters $ \mathbf{\Theta}$.

There exist a number of masks designed for speech enhancement,
such as ideal binary mask (IBM) \cite{hu2001speech}, IRM \cite{hummersone2014ideal}, TBM \cite{gonzalez2014mask}, spectral magnitude mask (SMM) \cite{wang2014training} and phase-sensitive mask (PSM) \cite{erdogan2015phase}.
While IBM uses a hard binary label on each T-F unit,
IRM can be viewed as a soft version of IBM, which is defined as
\begin{equation}\label{Eq4}
IRM_{t,f} = {\left(\frac{{X_{t,f}^2}}{{X_{t,f}^2}+{N_{t,f}^2}}\right)}^\beta,
\end{equation}
where ${t}$ and ${f}$ denote the time and frequency indices, respectively,
$ X$ and $N$ are the spectrograms of the target clean speech and noise, respectively.
Accordingly, ${X_{t,f}^2}$ and $ {N_{t,f}^2}$ are the speech and noise energy of the $(t,f)$-th T-F unit, respectively.
The tunable parameter $\beta$ scaling the mask is typically chosen to 0.5.
Intuitively, IRM is a soft mask defined based on the speech and noise energy of each T-F unit.

In comparison, IBM and TBM are complementary with IRM as they are hard binary masks.
In order to maximize the complementarity in comparison with IRM, our fusion method considers a variant of the TBM defined only based on the target clean speech as
\begin{equation}\label{Eq2}
TBM_{t,f} = \left\{\!\begin{array}{cl}\! {1,}&{{\text{if  }}    X_{t,f} > \tau_{f}}\\ {0,}&{{\text{otherwise}}} \end{array} \right.,
\end{equation}
with
\begin{equation}\label{Eq3}
\tau_{f}=\frac{1}{M} \sum_{t=1}^{M}X_{t,f}.
\end{equation}
This definition of TBM is in fact a binary classification of each T-F unit based on the target speech energy.

\section{Proposed Method}
\label{sec:pagestyle}

We utilize multi-target learning to simultaneously estimate two complementary masks, IRM and TBM.
Meanwhile, mask fusion is adopted to make full use of the complementary information from the two masks in order to achieve better performance than single mask based methods.

\subsection{Multi-Target Learning}

\begin{figure}[t]
	\centering
	\includegraphics[width=7.5cm]{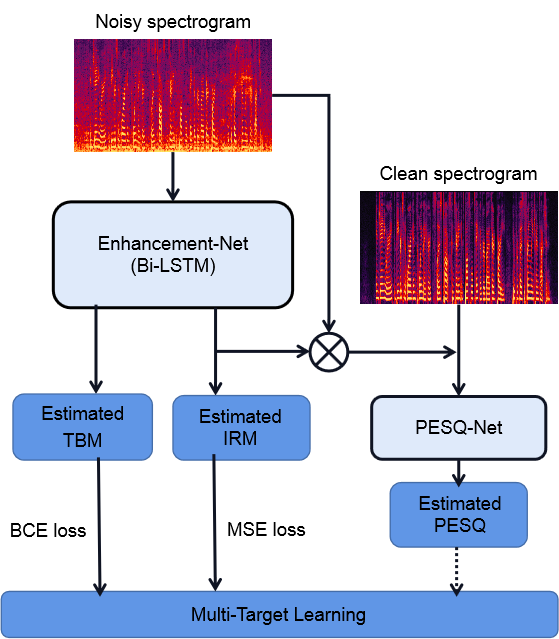}
	\caption{Overall architecture of the proposed multi-mask fusion model.
		The overall loss consists of a BCE loss of TBM and an MSE loss of IRM, with a perceptual loss being optional.}
	\label{fig:our network}
\end{figure}

Multi-target learning has been well demonstrated to be effective in enhancing the generalization ability of DNN models.
The overall architecture of the proposed model for speech enhancement is illustrated in Figure \ref{fig:our network}.
The network is designed to estimate the TBM and IRM from the noisy spectrogram,
which has two bi-directional long short-term memory (Bi-LSTM) layers.
Alternatively, the Bi-LSTM layers can be replaced with LSTM layers to make the system causal
and enabling streaming data processing.
In addition, a perceptual loss can be optionally added to further improve the performance.
Recently, perceptual loss has shown significant effectiveness for speech enhancement \cite{fu2019metricGAN,8902088,fu2018end,koizumi2018dnn}.

Accordingly, the overall loss consists of a loss for TBM, a loss for IRM, and optionally a perceptual loss,
which given by
\begin{equation}\label{Eq5}
\mathcal{L}=\mathcal{L}_{IRM}+{\alpha} \mathcal{L}_{TBM}+{\lambda} \mathcal{L}_{per},
\end{equation}
where $\alpha > 0$ and ${\lambda}>0$ are penalty parameters to balance between the three losses.
The IRM loss $\mathcal{L}_{IRM}$ is the MSE of IRM estimation as
\begin{equation}\label{Eq6}
{\mathcal{L}_{IRM}} = \sum\limits_{t,f} {{{\left( {\widehat{IRM}_{t,f} - IRM_{t,f}} \right)}^2}},
\end{equation}
where $ {\widehat{IRM}}_{t,f}$ is the estimated IRM by the network.
There exists another way to learn the target mask,
which is called signal approximation \cite{weninger2015speech,7032183}.
It trains a ratio mask estimator that minimizes the difference
between the spectrogram of clean speech and that of estimated speech,
which has been tested in our model but shown no performance improvement.
The TBM loss $\mathcal{L}_{TBM}$ is the binary cross entropy (BCE) loss of TBM estimation as
\begin{equation}\label{Eq7}
\begin{aligned}
{L}_{TBM} = &-\sum\limits_{t,f}TBM_{t,f}\cdot log({\widehat{TBM}}_{t,f})+\\
&~~~~~~~~(1-TBM_{t,f})\cdot log(1-{\widehat{TBM}}_{t,f}),
\end{aligned}
\end{equation}
where ${\widehat{TBM}}_{t,f}$ is the TBM estimated by the network.
For TBM estimation, which is a binary classification task of T-F bins, the BCE loss is more suitable than MSE.

The perceptual loss $ \mathcal{L}_{per}$ is based on an approximate PESQ estimation function,
for which a network is used to estimate PESQ \cite{fu2019metricGAN} since the standard PESQ function is non-differentiable.
Specifically, $\mathcal{L}_{per}$ is given by
\begin{equation}\label{Eq8}
\mathcal{L}_{per}=1-Q\left(Y_{t,f}\odot \widehat{IRM}_{t,f}, X_{t,f}\right),
\end{equation}
where $Y_{t,f}$ is the noisy spectrogram, and $Q$ is the PESQ estimation network, referred to as PESQ-Net.
Similar to \cite{8902088}, the PESQ-Net $Q$ is pre-trained beforehand and fixed in the procedure of training the mask estimation network.
The PESQ-Net is only used for training.

Note that, the perceptual loss is optional in our method.
We consider it in this paper in an order to show that,
when incorporated into the sophisticated perceptual training framework,
the proposed method can also yield performance improvement over existing perceptual training based methods, as will be shown later in Section 4.

\begin{figure}[t]
	\centering
	\includegraphics[width=7.5cm]{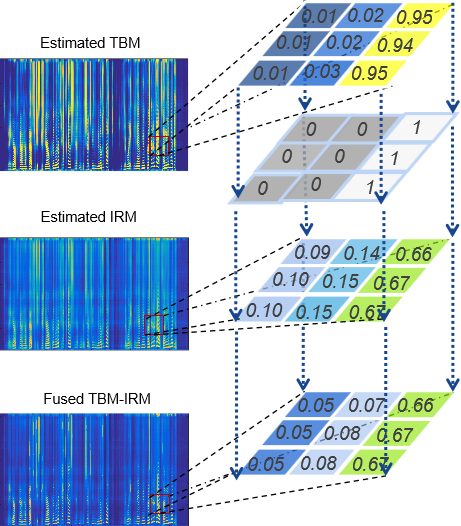}
	\caption{Illustration of mask fusion of IRM and TBM. The estimated TBM is first binarized and then fused with the estimated IRM to obtain a refined soft mask. The fused mask is finally used to mask the noisy spectrogram to suppress noise.}
	\label{fig:masks fusion}
\end{figure}

\subsection{Mask Fusion}

While the multi-target learning strategy aims to enhance the generalization ability of the mask estimation network,
the mask fusion strategy aims to make full use of the complementary information of the two different masks IRM and TBM.
Figure \ref{fig:masks fusion} illustrates the mask fusion strategy.
The basic idea is simply that, the T-F bins not dominated by speech energy according to estimated TBM should be weakened,
while the T-F bins dominated by speech energy according to estimated TBM shall remain unchanged.
The goal is to obtain a refined mask that is more robust against noise.
Specifically, the fused mask is computed as
\begin{equation}\label{Eq9}
MF_{t,f} = \left\{\!\begin{array}{cl}\! {{\widehat{IRM}}_{t,f},}&{{\text{if~  }}    {\widehat{TBM}}_{t,f} > \delta}\\ {\gamma\cdot{\widehat{IRM}}_{t,f} ,}&{{\text{otherwise}}} \end{array} \right.,
\end{equation}
where $0\!<\!\delta\!<\!1$ is a threshold parameter for the binarization of the estimated TBM,
$0\!\leq\!\gamma\!\leq\!1$ is the scale for weakening the T-F bins not dominated by speech.
When $\gamma=1$, it holds $MF_{t,f}\!=\!\widehat{IRM}_{t,f}$,
in this case the fused mask degenerates to the estimated IRM.
When $\gamma=0$, it is equivalent to fuse the binarized TBM with IRM directly by element-wise product.
Intensive experiments show that using a very small value of $\gamma$, e.g., $\gamma=0$,
would lead to degradation of speech perceptual quality caused by excessive suppression.

%\begin{figure}[t]
%	\centering
%	\includegraphics[width=7.5cm]{figure3.png}
%	\caption{Illustration of mask fusion of IRM and TBM. The estimated TBM is first binarized and then fused with the estimated IRM to obtain a refined soft mask. The fused mask is finally used to mask the noisy spectrogram to suppress noise.}
%	\label{fig:masks fusion}
%\end{figure}

\section{Experiments}

\subsection{Data Corpus}

We conduct experimental evaluation of the proposed method on the CHiME-4 challenge \cite{vincent2017analysis},
of which the data is collected in the scenarios where a person is talking to a mobile tablet device equipped with 6 microphones in a variety of adverse environments.
There are four noise conditions, caf\'{e} (CAF), street junction (STR), public transport (BUS), and pedestrian area (PED).
Simulated data of noisy speech are provided for each condition and is constructed by mixing clean utterances of the WSJ0 corpus \cite{garofalo2007csr} with environmental noise recordings using the method in \cite{vincent2007oracle}. But real data is recorded in real noisy environments uttered by actual talkers. Each recorded data has 6 channels. In this paper, we focus on the single channel case and only the data from channel-1 is used.

For the experiment on enhancement, all data is simulated in order to have clean speech as reference for training and testing. Specifically, the training set is generated by caf\'{e} noise, pedestrian noise, street noise and clean utterances. There are 5420 utterances used from 83 speakers, which aims to make our trained DNN speaker-independent. The development set are simulated with 1640 utterances.
For evaluation set, the test speech is taken from four speakers which are not seen during training or validation. The test noise is CAF, PED and STR but extracted from different files. And the bus noise is also used to generate the test noisy data to perform a test with even an unseen noise type.
Besides, we simulate four SNR conditions from -5dB to 10dB with a step size of 5dB for test to verify our method SNR-independent. As for the ASR experiment, both real and simulated data are used to test and we focus on the performance on the real test set.

\subsection{Implementation Details}

In our task, speech waveform is sampled at 16 kHz and the frame length is set to 512 samples (32 msec) with a frame shift of 256 samples (16 msec).
STFT analysis is used to compute the DFT of each overlapping windowed frame with hamming window.
And inverse-STFT is used to transform the T-F domain enhanced spectrogram into time-domain waveform.
Noisy spectrogram of each utterance with size of $N \times 257$ is used as the input of network.
Meanwhile, the reference IRM and TBM with the same size are used as the training labels.
The mask estimation network has four hidden layers, two Bi-LSTM layers of size $200$ and two dense layers of size $300$.
Two output layers activated by sigmoid with size $257$ are used to output the IRM and TBM, respectively.
First, we train the network with the loss function (\ref{Eq5}), in which $\alpha$ and $\lambda$ are set to 0.1 and 0, respectively.
That is the perceptual loss is not used.

Then, the perceptual loss using the PESQ-net is added to train the network again with parameter $\lambda=10$.
The PESQ-net is the same as that in \cite{8902088} and fixed in training the mask estimation network.
The ADAM algorithm is used to optimize the network, and the test performance in the development set is used to decide whether to update the trained model after each training epoch.
Besides, the parameters $ \delta$ and $ \gamma$ are selected based on the performance in the development set after finishing the training of mask estimation network.

\subsection{Experiment on Enhancement}

The performance of speech enhancement is evaluated in terms of PESQ.
The compared methods include:

\hangindent 2.5em
a) TBM: the masking-based method only using TBM;

\hangindent 2.5em
b) IRM: the masking-based method only using IRM;

\hangindent 2.5em
c) MTL1-TBM: the proposed method using multi-target learning but only using TBM without mask fusion;

\hangindent 2.5em
d) MTL1-IRM: the proposed method using multi-target learning but only using IRM without mask fusion;

\hangindent 2.5em
e) MTL1-FUS: the proposed fusion method;

\hangindent 2.5em
f) MTL2-TBM: the proposed method using multi-target learning and perceptual loss, but only using TBM without mask fusion;

\hangindent 2.5em
g) MTL2-IRM: the proposed method using multi-target learning and perceptual loss, but only using IRM without mask fusion;

\hangindent 2.5em
h) MTL2-FUS: the proposed fusion method additionally using perceptual loss;

\hangindent 2.5em
i) MetricGAN: the state-of-the-art perceptual loss based method \cite{fu2019metricGAN}.

Note that, all these compared methods use the same mask estimation network as described in Section 4.2
except for the output layers, e.g., TBM, IRM and MetricGAN use a single output layer, while the variants of the proposed method use two. Our method casues a little increase of network complexity because the PESQ-net will not be used for test and only an extra output layer is added. Specifically, the parameters of network are increased from 1.98M to 2.06M by our method in the experiment. 

\begin{table}[!t]\footnotesize
	\caption{PESQ score of different methods for seen noise types (CAF, PED and STR) on the simulated test set at SNR=5dB.  }
	\label{tab:result of pesq}
	\centering
	\begin{tabular}{ c c c c c }
		\hline
		\textbf{ }&{CAF}&{PED}&{STR}&{AVG} \\
		\hline
		\specialrule{0em}{1pt}{1pt}
		Noisy                  &$1.855$      &$1.804$      &$1.960$      &$1.873$ \\
		\specialrule{0em}{1pt}{1pt}
		\hline
		\specialrule{0em}{1pt}{1pt}
		TBM                  &$2.332$      &$2.319$      &$2.463$      &$2.371$ \\
		\specialrule{0em}{1pt}{1pt}
		IRM                   &$2.393$      &$2.382$      &$2.510$      &$2.428$ \\
		\specialrule{0em}{1pt}{1pt}
		MTL1-TBM                   &$2.347$      &$2.331$      &$2.472$      &$2.383$ \\
		\specialrule{0em}{1pt}{1pt}
		MTL1-IRM                   &$2.399$      &$2.382$      &$2.513$      &$2.431$ \\
		\specialrule{0em}{1pt}{1pt}
		MTL1-FUS                &$\bm{2.498}$      &$\bm{2.482}$      &$\bm{2.612}$      &$\bm{2.531}$ \\
		\specialrule{0em}{1pt}{1pt}
		\hline
		\specialrule{0em}{1pt}{1pt}
		MetricGAN      &$2.474$    &$2.487$    &$2.598$    &$2.520$ \\
		\specialrule{0em}{1pt}{1pt}
		MTL2-TBM                  &$2.367$      &$2.354$      &$2.486$      &$2.402$ \\
		\specialrule{0em}{1pt}{1pt}
		MTL2-IRM                  &$2.490$      &$2.497$      &$2.604$      &$2.530$ \\
		\specialrule{0em}{1pt}{1pt}
		MTL2-FUS               &$\bm{2.516}$      &$\bm{2.513}$      &$\bm{2.632}$      &$\bm{2.554}$ \\
		\specialrule{0em}{1pt}{1pt}
		\hline
	\end{tabular}
\end{table}

\begin{table}[!t]\footnotesize
	\caption{PESQ score of different methods for unseen noise types (BUS) on the simulated test set at all SNRs.  }
	\label{tab:bus result of pesq}
	\centering
	\begin{tabular}{ c c c c c c}
		\hline
		\textbf{ }&{-5dB}&{0dB}&{5dB}&{10dB}&{AVG} \\
		\hline
		\specialrule{0em}{1pt}{1pt}
		Noisy        &$1.445$          &$1.795$      &$2.160$      &$2.520$      &$1.980$ \\
		\specialrule{0em}{1pt}{1pt}
		\hline
		\specialrule{0em}{1pt}{1pt}
		TBM         &$1.682$         &$2.207$      &$2.628$      &$2.943$      &$2.365$ \\
		\specialrule{0em}{1pt}{1pt}
		IRM        &$1.837$           &$2.295$      &$2.678$      &$3.004$      &$2.454$ \\
		\specialrule{0em}{1pt}{1pt}
		MTL1-TBM      &$1.693$             &$2.223$      &$2.634$      &$2.929$      &$2.370$ \\
		\specialrule{0em}{1pt}{1pt}
		MTL1-IRM         &$1.846$          &$2.304$      &$2.690$      &$3.020$      &$2.465$ \\
		\specialrule{0em}{1pt}{1pt}
		MTL1-FUS       &$\bm{1.918}$         &$\bm{2.399}$      &$\bm{2.788}$      &$\bm{3.098}$      &$\bm{2.551}$ \\
		\specialrule{0em}{1pt}{1pt}
		\hline
		\specialrule{0em}{1pt}{1pt}
		MetricGAN    &$1.867$  &$2.359$    &$2.759$    &$3.093$    &$2.520$ \\
		\specialrule{0em}{1pt}{1pt}
		MTL2-TBM      &$1.681$            &$2.225$      &$2.649$      &$2.955$      &$2.378$ \\
		\specialrule{0em}{1pt}{1pt}
		MTL2-IRM       &$1.888$           &$2.373$      &$2.760$      &$3.083$      &$2.526$ \\
		\specialrule{0em}{1pt}{1pt}
		MTL2-FUS      &$\bm{1.932}$         &$\bm{2.410}$      &$\bm{2.790}$      &$\bm{3.104}$      &$\bm{2.559}$ \\
		\specialrule{0em}{1pt}{1pt}
		\hline
	\end{tabular}
\end{table}

Table \ref{tab:result of pesq} shows the PESQ score of the compared methods on the simulated test set of CHiME-4 for seen noise types at SNR=5dB. Meanwhile, Table \ref{tab:bus result of pesq} shows the PESQ score for unseen noise type at all SNRs.
We can draw the following results from Table \ref{tab:result of pesq} and \ref{tab:bus result of pesq}.
First, IRM yields higher PESQ score than TBM, which implies that IRM is superior over TBM for speech perceptual quality.
Second, MTL1-IRM outperforms IRM, whilst MTL2-IRM outperforms MTL1-IRM in terms of PESQ.
Third, the proposed mask fusion can further significantly improve the performance,
which results in better performance of MTL1-FUS over MTL1-IRM and better performance of MTL2-FUS over MTL2-IRM.

The idea of multi-target learning can be used to learn multiple targets with increasing the complexity of the network slightly, but only have slight improvement in our framework. In comparison, the proposed mask fusion strategy can achieve significant improvement.
For example, even in the case without perceptual training, MTL1-FUS has better performance than the perceptual training based MetricGAN method.
When perceptual loss is used, the proposed method can achieve further improvement,
e.g., MTL2-FUS achieved the highest PESQ score in all four conditions and all SNRs.
However, the advantage of MTL1-FUS over MTL1-IRM is more prominent than that of MTL2-FUS over MTL2-IRM.

Table \ref{tab:different se} and \ref{tab:different th} shows the PESQ comparison for different parameter of fusion. Table \ref{tab:different se} shows the different scale for weakening the T-F bins not dominated by speech when the threshold is fixed. The optimal performance is obtained when $ \gamma$ is set to about 0.5. Meanwhile, the excessive weakening of T-F bins such as $ \gamma=0.1 $ will lead to a drop of PESQ scores and serious speech distortion. When $ \gamma=1 $, it will not weaken the T-F bins and is equivalent to no fusion. Table \ref{tab:different th} shows the different threshold for the binarization of the estimated TBM when the scale is fixed. The optimal performance is obtained when $ \delta$ is set to about 0.9, which is different from the scale. The great threshold means there are more T-F bins weakened. However, it makes no sense that $ \delta=1$ is set to 1 because all T-F bins are weakened. Besides, if $ \delta=0$ there will be no T-F bins weakened, which is equivalent to $ \gamma=1 $ and has the same result.
\begin{table}[!t]\footnotesize
	\caption{PESQ comparison of different scale $ \gamma$ for MTL1-FUS when $ \delta=0.5$.  }
	\label{tab:different se}
	\centering
\begin{tabular}{ c c c c c c}
	\hline
	\textbf{$\gamma$}&{-5dB}&{0dB}&{5dB}&{10dB}&{AVG} \\
	\hline
	\specialrule{0em}{1pt}{1pt}
	$0$        &$1.775$          &$2.256$      &$2.600$      &$2.824$      &$2.364$ \\
	\specialrule{0em}{1pt}{1pt}
	$0.1$         &$1.775$         &$2.257$      &$2.604$      &$2.835$      &$2.368$ \\
	\specialrule{0em}{1pt}{1pt}
	$0.2$        &$1.791$           &$2.289$      &$2.674$      &$2.958$      &$2.428$ \\
	\specialrule{0em}{1pt}{1pt}
	$0.3$        &$1.825$           &$2.335$      &$2.744$      &$3.059$      &$2.491$ \\
	\specialrule{0em}{1pt}{1pt}
	$0.4$      &$1.853$             &$2.364$      &$2.778$      &$3.104$      &$2.525$ \\
	\specialrule{0em}{1pt}{1pt}
	$0.5$         &$1.869$          &$\bm{2.373}$      &$\bm{2.784}$      &$\bm{3.115}$      &$\bm{2.535}$ \\
	\specialrule{0em}{1pt}{1pt}
	$0.6$       &$\bm{1.875}$         &$2.370$      &$2.775$      &$3.107$      &$2.532$ \\
	\specialrule{0em}{1pt}{1pt}
	$0.7$    &$1.874$  &$2.358$    &$2.758$    &$3.089$    &$2.520$ \\
	\specialrule{0em}{1pt}{1pt}
	$0.8$      &$1.868$            &$2.342$      &$2.736$      &$3.067$      &$2.503$ \\
	\specialrule{0em}{1pt}{1pt}
	$0.9$       &$1.858$           &$2.324$      &$2.713$      &$3.044$      &$2.485$ \\
	\specialrule{0em}{1pt}{1pt}
	$1$      &$1.846$         &$2.304$      &$2.690$      &$3.020$      &$2.465$ \\
	\specialrule{0em}{1pt}{1pt}
	\hline
\end{tabular}
\end{table}

\begin{table}[!t]\footnotesize
	\caption{PESQ comparison of different threshold $ \delta$ for MTL1-FUS when $ \gamma=0.5$.  }
	\label{tab:different th}
	\centering
	\begin{tabular}{ c c c c c c}
		\hline
		\textbf{$\delta$}&{-5dB}&{0dB}&{5dB}&{10dB}&{AVG} \\
		\hline
		\specialrule{0em}{1pt}{1pt}
		$0$       &$1.846$         &$2.304$      &$2.690$      &$3.020$      &$2.465$ \\
		\specialrule{0em}{1pt}{1pt}
		$0.1$         &$1.828$         &$2.321$      &$2.741$      &$3.098$      &$2.497$ \\
		\specialrule{0em}{1pt}{1pt}
		$0.2$        &$1.836$           &$2.337$      &$2.760$      &$3.108$      &$2.510$ \\
		\specialrule{0em}{1pt}{1pt}
		$0.3$      &$1.845$             &$2.351$      &$2.770$      &$3.113$      &$2.520$ \\
		\specialrule{0em}{1pt}{1pt}
		$0.4$         &$1.858$          &$2.363$      &$2.778$      &$\bm{3.115}$      &$2.529$ \\
		\specialrule{0em}{1pt}{1pt}
		$0.5$       &$1.869$         &$2.373$      &$2.784$      &$\bm{3.115}$      &$2.535$ \\
		\specialrule{0em}{1pt}{1pt}
		$0.6$    &$1.881$ 		 &$2.382$    &$2.788$    &$\bm{3.115}$    &$2.542$ \\
		\specialrule{0em}{1pt}{1pt}
		$0.7$      &$1.894$            &$2.390$      &$\bm{2.791}$      &$3.111$      &$2.546$ \\
		\specialrule{0em}{1pt}{1pt}
		$0.8$       &$1.909$           &$2.395$      &$2.790$      &$3.105$      &$\bm{2.550}$ \\
		\specialrule{0em}{1pt}{1pt}
		$0.9$      &$\bm{1.921}$         &$\bm{2.399}$      &$2.784$      &$3.094$      &$\bm{2.550}$ \\
		\specialrule{0em}{1pt}{1pt}
		$1$      &$1.849$         &$2.301$      &$2.685$      &$3.016$      &$2.463$ \\
		\specialrule{0em}{1pt}{1pt}
		\hline
	\end{tabular}
\end{table}

\begin{figure}[t]
	\centering
	\includegraphics[width=7.5cm]{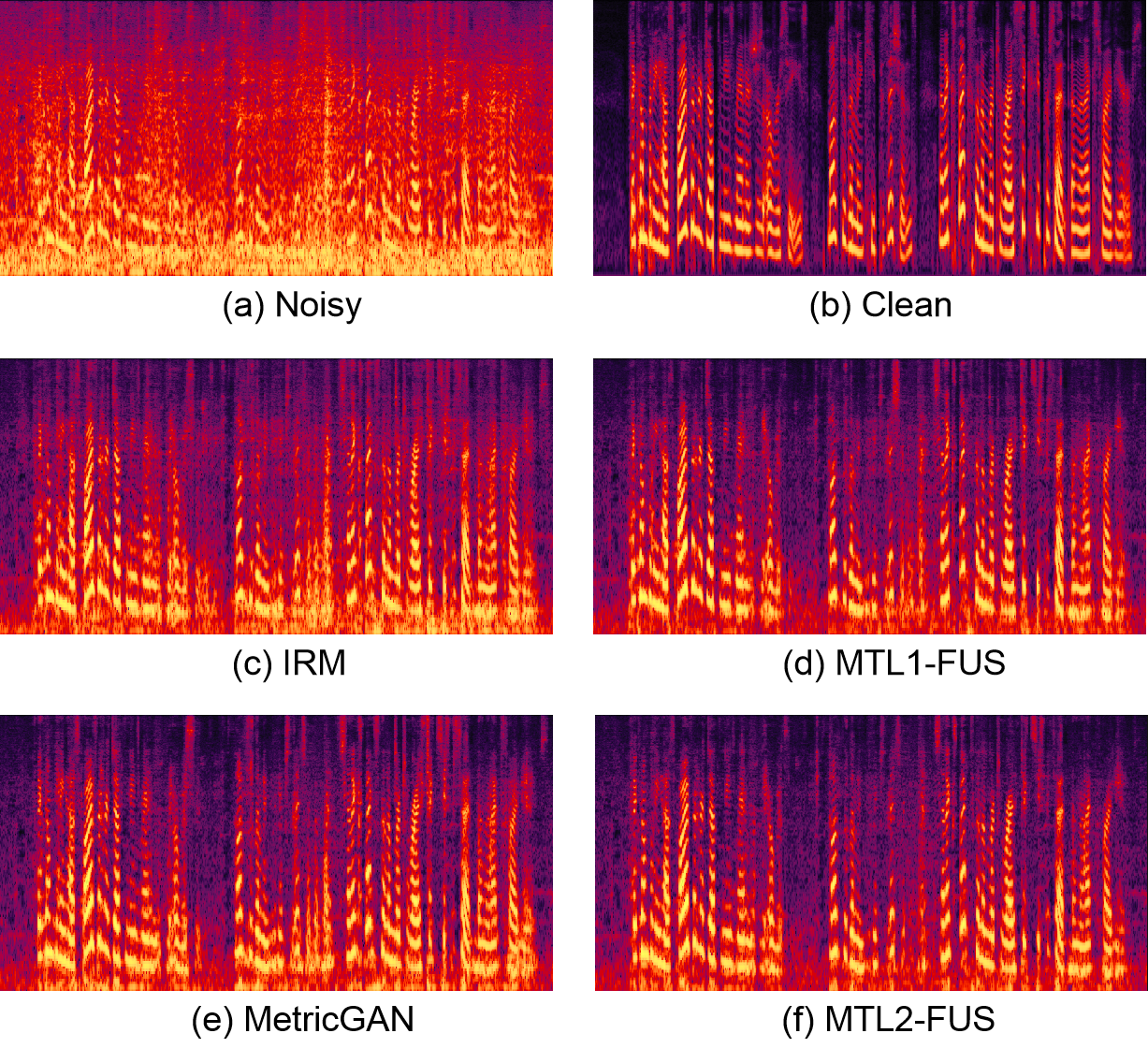}
	\caption{Magnitude spectrograms of the IRM, MTL1-FUS, MetricGAN, and MTL2-FUS methods.} 
	\label{fig:spec}
\end{figure}

Figure \ref{fig:spec} compares the enhanced magnitude spectrograms of an utterance.
The noisy spectrogram contains high noise at the entire spectrogram, especially at low frequency bins.
All the DNN based approaches can achieve good performance for speech enhancement.
However, there still exits obvious noise at low frequency bins of the spectrogram enhanced by IRM,
as shown in Figure \ref{fig:spec} (c), which would significantly degrade the speech perceptual quality.
Mask fusion can effectively reduce the noise at low frequency bins, as shown in Figure \ref{fig:spec} (d),
which results in significant improvement in speech perceptual quality (hence higher PESQ score).
MetricGAN and MTL2-FUS can recover the target speech spectrogram better than IRM.
Some audio samples is offered at https://github.com/lc-zhou/mask-fusion/tree/main/audio\_sample.

\subsection{Experiment on ASR}
The ASR system is trained on the speech recognition toolkit Kaldi \cite{Povey_ASRU2011} for the back-end configurations. There are HMM-GMM and TDNN acoustic models used in our experiment and the decoding is based on N-gram language models. The performance of ASR is evaluated in terms of word error rate (WER) and enhanced speech is directly fed into ASR system. Table \ref{tab:result of dt et} has shown the WER comparison for three front-end methods on the different set. IRM, MTL2-IRM and MTL2-FUS have been explained in Section 4.3 and are selected for test.
As the front-end, MTL2-IRM has obtained the better result than other methods at both back-end configurations. Particularly, there are obvious improvement between MTL2-IRM and IRM. Multi-target learning again can achieve improvement compared with traditional single mask based methods, but mask fusion no longer leads to improvement for WER.
Tabel \ref{tab:result of wer} shows the WER for different noisy types on the real test set. The methods have the best performance for street junction noise but not good for bus noise.

\begin{table}[!t]\footnotesize
	\caption{WER comparison of three front-end methods on the different set.  }
	\label{tab:result of dt et}
	\centering
	\begin{tabular}{ c c c c c }
		\hline
		\textbf{ }&{dt\_simu}&{dt\_real}&{et\_simu}&{et\_real} \\
		\hline
		\specialrule{0em}{1pt}{1pt}
		IRM-GMM        &$28.45$          &$28.64$      &$35.73$      &$41.71$       \\
		\specialrule{0em}{1pt}{1pt}
		MTL2-IRM-GMM       &$\bm{18.78}$       &$17.90$      &$\bm{25.85}$   &$\bm{29.63}$     \\
		\specialrule{0em}{1pt}{1pt}
		MTL2-FUS-GMM      &$18.96$   &$\bm{17.74}$   &$27.03$   &$30.11$   \\
		\specialrule{0em}{1pt}{1pt}
		\hline
		\specialrule{0em}{1pt}{1pt}
		IRM-TDNN    &$17.89$  &$15.94$    &$26.77$    &$29.14$   \\
		\specialrule{0em}{1pt}{1pt}
		MTL2-IRM-TDNN     &$\bm{12.66}$   &$\bm{10.67}$      &$\bm{20.60}$      &$\bm{22.26}$  \\
		\specialrule{0em}{1pt}{1pt}
		MTL2-FUS-TDNN     &$14.44$   &$12.31$      &$23.55$      &$25.05$       \\
		\specialrule{0em}{1pt}{1pt}
		\hline
	\end{tabular}
\end{table}

\begin{table}[!t]\footnotesize
	\caption{WER comparison of different noisy types on the real test set.  }
	\label{tab:result of wer}
	\centering
	\begin{tabular}{ c c c c c c }
		\hline
		\textbf{ }&{BUS}&{CAF}&{PED}&{STR}&{AVG} \\
		\hline
		\specialrule{0em}{1pt}{1pt}
		IRM-GMM        &$53.56$          &$45.57$      &$39.14$      &$28.58$      &$41.71$ \\
		\specialrule{0em}{1pt}{1pt}
		MTL2-IRM-GMM       &$\bm{40.19}$       &$31.90$      &$\bm{26.10}$   &$\bm{20.32}$      &$\bm{29.63}$ \\
		\specialrule{0em}{1pt}{1pt}
		MTL2-FUS-GMM      &$41.11$   &$\bm{31.81}$   &$26.74$   &$20.81$      &$30.11$ \\
		\specialrule{0em}{1pt}{1pt}
		\hline
		\specialrule{0em}{1pt}{1pt}
		IRM-TDNN    &$41.32$  &$29.98$    &$26.46$    &$18.81$    &$29.14$ \\
		\specialrule{0em}{1pt}{1pt}
		MTL2-IRM-TDNN     &$\bm{34.90}$   &$\bm{21.18}$      &$\bm{19.02}$      &$\bm{13.93}$      &$\bm{22.26}$ \\
		\specialrule{0em}{1pt}{1pt}
		MTL2-FUS-TDNN     &$39.63$   &$24.58$      &$20.16$      &$15.84$      &$25.05$ \\
		\specialrule{0em}{1pt}{1pt}
		\hline
	\end{tabular}
\end{table}

\section{Conclusions}

A multi-mask fusion method has been proposed for speech enhancement,
which adopts multi-target learning to simultaneously estimate two complementary masks, namely TBM and IRM, and fuses them to achieve better enhancement performance.
Experimental results demonstrated that the proposed fusion method can achieve the PESQ improvement over traditional single mask based methods.
Furthermore, when perceptual loss is incorporated, the proposed method can achieve further improvement
and outperforms the state-of-the-art perceptual loss based method.
Besides, as the fornt-end the proposed method can also achieve more reduction of WER than traditional single mask based methods.
In future study, we will explore other targets related to speech enhancement and other methods that can improve the speech perceptual quality and reduce the recognition error rate.

\bibliographystyle{IEEEbib}
\bibliography{mybib}

\end{document}